\documentclass{iau}
\usepackage{graphicx}

\newfont{\myfont}{cmmib10}

\newcommand{\bomega}{\hbox{\myfont \symbol{33} }}

\def\GJ{{\rm GJ}}
\def\bi{\bf}

\def\div{{\rm div\,}}
\def\curl{{\rm curl\,}}

\def\be{\begin{equation}}
\def\ee{\end{equation}}
\def\bea{\begin{eqnarray}}
\def\eea{\end{eqnarray}}

\title[IAUS291.~~Pulsar electrodynamics revisited] 
{Pulsar electrodynamics revisited}

\author[D.~B.~Melrose \& R.~Yuen]   
{D.~B.~Melrose \and R.~Yuen}

\affiliation{SIfA, School of Physics, The University of Sydney\\ 
NSW 2006, Australia\\ email: {\tt melrose@physics.usyd.edu.au} }

\pubyear{2012}
\volume{291}  
\pagerange{}
\jname{\mbox{Neutron Stars and Pulsars: Challenges and Opportunities after 80 years}}
\editors{J.~van Leeuwen, ed.} 
\begin{document}

\maketitle

\begin{abstract}
The inductive electric field is unjustifiably neglected in most  models for pulsar electrodynamics; it cannot be screened by the magnetospheric plasma, and it is not small in comparison with the corotation electric field. The perpendicular component of the inductive electric field implies a drift motion that is inconsistent with corotation at any angular velocity. Some implications of the inductive electric field and the associated drift motion are discussed.
\end{abstract}

\firstsection 
\section{Introduction}

Pulsar electrodynamics, as developed four decades ago, includes some unresolved inconsistencies. The vacuum dipole model (VDM) is used for some purposes and the corotating magnetosphere model (CMM) is used for other purposes. Recent observational evidence for a link between magnetospheric phenomena (nulling, mode changing, timing noise) and the rate of slowing down (\cite{Ketal06,Letal10}) seems to require an explanation that involves features of both the VDM and the CMM. It is timely to consider how the incompatibility between the VDM and the CMM can be resolved.

In the VDM, the pulsar is modeled as an obliquely (angle $\alpha$) rotating magnetic dipole in vacuo. The energy and angular momentum are carried off by magnetic dipole radiation with a frequency equal to the rotation frequency of the star. The VDM is used to estimate the surface magnetic field, $B_*\sin\alpha\propto(P{\dot P})^{1/2}$, and the pulsar age, $P/2{\dot P}$, from the observed period, $P$, and period derivative, ${\dot P}$. In the CMM, it is postulated that the magnetosphere is corotating with the star, requiring that the electric field be the corotation field, ${\bi E}_{\rm cor}$, whose divergence determines the Goldreich-Julian charge density, $\rho_\GJ$ (\cite{GJ69}). As conventionally formulated, the CMM is quasi-electrostatic and quasi-stationary, due to assuming either an aligned rotator ($\alpha=0$), or that the magnetosphere is stationary in the corotating frame. The CMM is the basis for most detailed models of the electrodynamics, in which $\rho_\GJ$ can be provided by charges of one sign just above the stellar surface in the polar cap, but requires an additional source of charge, through pair creation in vacuum gaps (\cite{RS75}), at greater heights. Both the VDM and CMM are fatally flawed as stand-alone models. 

Two unacceptable features of the VDM are that 1)  the magnetic dipole radiation cannot escape because its frequency is below the plasma frequency (in the magnetosphere, the wind and the ISM), and 2) there is no plasma to emit the observed radiation. Two unacceptable features of the CMM are that 1) the logical consequence of the assumptions made in the CMM is a completely charge-separated (`domed') electrosphere (\cite{M04}),  and 2) quasi-stationary models are unstable to growth of large-amplitude electric oscillations when subject to temporal perturbations (\cite{Letal05,BT07,T10}). 

The inductive electric field, ${\bi E}_{\rm ind}$, and the displacement current, $\varepsilon_0\partial{\bi E}_{\rm ind}/\partial t$, are essential features of the VDM, and are  ignored in the CMM. Including ${\bi E}_{\rm ind}$ involves a substantial rethinking of pulsar electrodynamics. A self-consistent model needs to start from the VDM and incorporate the response (charge and current densities) of the plasma that modify the vacuum field, leading to a self-consistent field. We outline an EM/plasma model that includes this response, and identify some implications of the model.

\section{Vacuum dipole and Corotating models}

\noindent
{\bf VDM}: The exact form of the magnetic field for a rotating point dipole, ${\bf m}$, is
\be{\bf B}(t,{\bf x})={\mu_0\over4\pi}
\left[{3{\bf x}\,{\bf x}\cdot{\bf m}-r^2{\bf m}\over r^5}+{3{\bf x}\,{\bf x}\cdot{\dot{\bf m}}-r^2{\dot{\bf m}}\over r^4c}
+{{\bf x}\times({\bf x}\times{\ddot{\bf m}})\over r^3c^2}
\right],
\label{VDM1}
\ee
with $\dot{\bf m}=\bomega\times {\bf m}$, $\ddot{\bf m}=\bomega\times(\bomega\times {\bf m})$. The electric field is
\be{\bf E}(t,{\bf x})={\mu_0\over4\pi}\left[
{{\bf x}\times{\dot{\bf m}}\over r^3}
+{{\bf x}\times{\ddot{\bf m}}\over r^2c}\right].
\label{VDM2}
\ee
These fields are included in the Deutsch (1955) model, which also includes a quadrupolar electrostatic field, due to ${\bi E}_{\rm cor}$ inside the star implying a surface charge on the star.

Using labels dip = dipolar,  ind = inductive, rad = radiative, with $ r_L=c/\omega$ the light cylinder radius, the magnetic field has terms ${\bf B}_{\rm dip}\propto1/r^3$,  ${\bf B}_{\rm ind}\propto1/r_Lr^2$, ${\bf B}_{\rm rad}\propto1/r_L^2r$, and the electric field has ${\bf E}_{\rm ind}\propto1/r_Lr^2$, ${\bf E}_{\rm rad}\propto1/r_L^2r$.  Only the dipole term, which dominates at $r\ll r_L$, and the radiative terms, which dominate at $r\gg r_L$, are included in the conventional VDM. Here we emphasize the role of the inductive terms.

\noindent
{\bf CMM}: The corotation electric field is
\be 
{\bf E}_{\rm cor}=-(\bomega\times{\bi x})\times{\bi B},
\qquad
\rho_{\rm GJ}=\varepsilon_0\div{\bf E}_{\rm cor},
\label{CMM1}
\ee 
usually approximated by $\rho_{\rm GJ}=-2\varepsilon_0\bomega\cdot{\bf B}_{\rm dip}$. Thus, in the CMM the electric field is postulated, and Maxwell's equations are used only to infer consequences (notably $\rho_{\rm GJ}$) of this postulate. In contrast, in the VDM the rotating magnetic field is postulated, and Maxell's equations (with $\rho=0$, ${\bi J}=0$) are solved for ${\bi E}_{\rm ind}$.

A prerequisite for assuming (\ref{CMM1}) and ignoring (\ref{VDM2}) is that the inductive field is either screened or is negligible. The latter is not the case: the ratio $|{\bf E}_{\rm ind}|/|{\bf E}_{\rm cor}|\propto\sin\alpha$ is of order unity independent of $r$. Screening is ineffective for the perpendicular component of  ${\bf E}_{\rm ind}$, and its neglect is not justified  (except for $\alpha\to0$). When ${\bf E}_{\rm ind}$ is included, the corotation postulate becomes untenable.

\section{EM/plasma model}

\noindent 
{\bf Including plasma in the VDM}: An EM/plasma model may be identified by starting from the VDM and including the response of the plasma to the inductive electric field. This field is of low frequency, compared with other frequencies in the plasma. The simplest useful plasma model is the low-frequency limit of the response of a cold electron-positron gas.  The cold-plasma model is used in a related application to auroral electron acceleration (\cite{SL06}), and although it is an oversimplification, e.g., the neglect of relativistic effects, an argument (given below) based on the polarization drift supports the use of the cold plasma approximation.  The response to the components, ${\bi E}_{{\rm ind}\,\perp}$ and $E_{{\rm ind}\,\parallel}$,  perpendicular and parallel to the magnetic field, depends on the Alfv\'en speed, $v_A$, and the plasma frequency, $\omega_p$, respectively. We also include the effect of a conductivity tensor with components $\sigma_\perp$, $\sigma_\parallel$. 

After inverting the (temporal) Fourier transform, the relations implied by the cold-plasma response are
\be
{\bi J}_\perp={c^2\over v_A^2}{\bi J}_{\rm disp\,\perp}+\sigma_\perp{\bi E}_\perp,
\qquad 
{\partial J_\parallel\over\partial t}=\varepsilon_0\omega_p^2E_\parallel
+\sigma_\parallel{\partial E_\parallel\over\partial t},
\label{EMP1}
\ee
where ${\bi J}_{\rm disp}=\varepsilon_0{\partial{\bi E}/\partial t}$ the displacement current. 

\noindent 
{\bf Oscillatory parallel response}: The parallel response of the plasma is strongly oscillatory, as has been identified in the pulsar literature (\cite{S71,Letal05,T10}). The resulting large-amplitude oscillations lead to pair creation, and the pairs can screen $E_{{\rm ind}\parallel}$.

\noindent 
{\bf Perpendicular response}: It is impossible to screen ${\bi E}_{{\rm ind}\perp}$ by charges. ${\bi E}_{{\rm ind}\perp}$ can be screened by currents if these lead to a changing magnetic field that balances $\partial{\bi B}/\partial t$. This may occur inside the star, but is ineffective in a pulsar magnetosphere.

Including the perpendicular response (\ref{EMP1}) in Maxwell's equations gives
\be
(\curl{\bi B})_\perp={1\over v_A^2}{\partial{\bi E}_\perp\over\partial t}
+{1\over c^2}{\partial{\bi E}_\perp\over\partial t}
+\mu_0\sigma_\perp{\bi E}_\perp.
\label{EMP2}
\ee
Due to the strong field and low density in a pulsar magnetosphere, $\sigma_\perp$ is small, and the final term in (\ref{EMP2}) can be neglected. (A collisional model for the conductivity gives $\sigma_\perp\propto\nu_e\to0$, $\sigma_\parallel\propto1/\nu_e\to\infty$ for collision frequency $\nu_e\to0$.) The strong field and low density also implies $v_A^2\gg c^2$. The term involving $1/v_A^2$ in (\ref{EMP2}) is then small compared with the term involving $1/c^2$. Thus the plasma current is negligible in comparison with the displacement current. It follows that plasma does not modify ${\bi E}_{{\rm ind}\perp}$ significantly from its vacuum value, where the displacement current balances $\curl{\bi B}$.

\noindent
{\bf Inductively induced drift}: 
A perpendicular electric field in a magnetized plasma induces an electric drift in which all particles participate, irrespective of their charge. The important physical effect that is neglected by ignoring ${\bi E}_{{\rm ind}\perp}$ is the inductive drift, ${\bi v}_{\rm ind}={\bi E}_\perp\times{\bi B}/B^2$, or ${\bi E}_{{\rm ind}\perp}=-{\bi v}_{\rm ind}\times{\bi B}$. This drift is of order the corotation speed times $\sin\alpha$, and has all components, $v_r,v_\theta,v_\phi$, nonzero (Melrose \& Yuen 2012).  The azimuthal component, $v_{\rm ind\,\phi}=2\omega r\sin\alpha\cos\theta_m/(1+3\cos^2\theta_m)$, where $\theta_m$ is the magnetic colatitude of the emission point, does not correspond to rotation at any angular speed (\cite{MY12}).

\noindent
{\bf Polarization drift}: 
The displacement current, due to the time-derivative of ${\bi E}_{{\rm ind}\perp}$, induces a polarization drift. Electrons and positrons drift in opposite directions, leading to the polarization current, which reproduces the term $(c^2/ v_A^2)({\bi J}_{\rm disp})_\perp$ in (\ref{EMP1}), with $v_A^2$ defined to include relativistic effects. This alternative interpretation of (\ref{EMP1}) justifies the use of the cold-plasma model.

\section{Implications of EM/plasma model}

We mention four possible implications of the inclusion of ${\bi E}_{\rm ind}$.

\noindent
{\bf Magnetosphere cannot be corotating}: Suppose that the inductive drift is superimposed on a corotation motion. The sign of $v_{\rm ind\,\phi}$ depends on the sign of $\cos\theta_m$, where $\theta_m$ is the magnetic colatitude. In a statistical sample of pulsars one expects half to have $\cos\theta_m>0$ and half to have $\cos\theta_m<0$. The implication is that, statistically, the source region in half of all pulsars would appear to be super-rotating, and half would appear to be sub-rotating.

\noindent
{\bf Subpulse drifting}: It may be that subpulse drifting is an observational consequence of this inductive drift. We are developing a detailed model; the first step is to identify the values of $\alpha$ and $\theta_m$ that an observer whose line of sight is at angle $\zeta$ can see as a function of the phase, $\chi$, and the radius, $r$, of the emission point.

\noindent
{\bf Definition of the polar cap}: Conventionally, the boundary of the polar cap region is defined by the last closed field line, and there is no mixing of plasma across this boundary. However, the nonzero radial and polar components of ${\bi v}_{\rm ind}$ imply that there is a plasma flow across the boundary, invalidating the concept of a well-defined boundary. This boundary is usually defined by retaining on ${\bi B}_{\rm dip}$, and ${\bi B}_{\rm ind}$ also needs to be included, for heights $r/r_L>0.1$ say (\cite{BS10}). Thus the closed field region can be a source of plasma for the polar cap region.

\noindent
{\bf Acceleration of $\gamma$-ray emitting particles}: Screening of $E_{{\rm ind}\parallel}$ should occur provided sufficient charges are available. $\gamma$-ray emission is attributed to acceleration by $E_\parallel$ in an outer (or slot) gap, where screening becomes ineffective due to charge starvation. The breakdown of screening causes $E_{\rm ind\,\parallel}$ to appear. Rather than acceleration being due to a putative electrostatic field across a gap (\cite{Tetal10}), it is due to $E_{{\rm ind}\parallel}$ becoming (partially) unscreened and available to accelerate charges.

\section{Conclusions}

A viable (EM/plasma) model for pulsar electrodynamics needs to be fully electromagnetic and to include the response of the plasma to the EM fields. The first steps in formulating such a model is to start from the exact form, (\ref{VDM1}) and (\ref{VDM2}), of the fields for the VDM, and to include the plasma response to these fields, e.g., in the form (\ref{EMP1}). 

The parallel plasma response is oscillatory, leading to pair creation (\cite{Letal05,BT07,T10}). The charges can screen $E_{{\rm ind}\parallel}$ in the inner magnetosphere. Breakdown of screening in the outer magnetosphere due to charge starvation plausibly accounts for acceleration of $\gamma$-ray emitting particles.

The perpendicular plasma response is ineffective in screening ${\bi E}_{{\rm ind}\perp}$, implying an inductive drift that has not been included in existing models. The presence of the inductive drift, ${\bi v}_{\rm ind}$, implies that the magnetosphere cannot be corotating, contrary to the central assumption in the CMM. The form of ${\bi v}_{\rm ind}$ also implies a flow across the last closed field line, allowing plasma transfer between the ``open'' and ``closed'' regions. The possible interpretation of subpulse drifting in terms of ${\bi v}_{\rm ind}$ is under investigation.

\end{document}